\documentclass[english]{revtex4-1}
\usepackage[T1]{fontenc}
\usepackage[latin9]{inputenc}
\usepackage{amsmath}
\usepackage{graphicx}
\usepackage{wasysym}

\makeatletter
\numberwithin{equation}{section}

\makeatother

\usepackage{babel}
\begin{document}

\title{Uniform analytic approximation of Wigner rotation matrices}

\author{Scott E. Hoffmann}
\email{scott.hoffmann@uqconnect.edu.au}

\address{School of Mathematics and Physics~\\
University of Queensland~\\
Brisbane QLD 4072~\\
Australia}
\begin{abstract}
We derive the leading asymptotic approximation, for low angle $\theta$,
of the Wigner rotation matrix elements $d_{m_{1}m_{2}}^{j}(\theta)$,
uniform in $j,m_{1}$ and $m_{2}.$ The result is in terms of a Bessel
function of integer order. We numerically investigate the error for
a variety of cases and find that the approximation can be useful over
a significant range of angles. This approximation has application
in the partial wave analysis of wavepacket scattering.
\end{abstract}
\maketitle

\section{Introduction}

The purpose of this paper is to derive an approximation for the Wigner
rotation matrices, $d_{m_{1}m_{2}}^{j}(\theta)$, as a function of
the angle $\theta$ and uniform in $j,m_{1}$ and $m_{2},$ for use
in analytic calculations.

There are several methods available for computing individual Wigner
rotation matrix elements to high precision. Wigner's series for the
matrix elements (equivalent to the terminating hypergeometric series
in Eq. (\ref{eq:9}) below) becomes, for large indices, a sum of very
large terms with alternating signs, exceeding the floating-point precision.
One of the alternative methods involves using recurrence relations
obeyed by the matrix elements \cite{Dachsel2006,Choi1999}. A precision
of 15 significant figures can be obtained. Another method involves
converting the sum into a Fourier series, which is better behaved
\cite{Feng2015,Tajima2015}. Fukushima \cite{Fukushima2016} presents
a method, using recurrence relations and extension of floating-point
exponents that can achieve 16 significant figures for very large values
of the indices.

The approximation presented here cannot obtain the very high precisions
of the methods just mentioned, as we will see below. However, it has
the advantage of giving the approximation as a function of the angle,
which can then be used in integrals.

The motivation for this work came from a recent paper by the author
\cite{Hoffmann2017a} on the scattering theory of wavepackets in a
Coulomb potential. The system considered was a single, nonrelativistic,
spinless particle, but the results presented here should have wider
applicability: to multiple particles, nonvanishing spins and relativistic
treatments \cite{Jacob1959,Macfarlane1962}.

It was necessary to transform the wavefunction from a basis of momentum
eigenvectors (with wavefunction $\Psi_{\mathrm{0}}(\boldsymbol{k})$)
to a basis of free eigenvectors of the magnitude of momentum, $k$,
and the familiar angular momentum quantum numbers, $l$ and $m$,
taking only integer values in this case (with wavefunction $\Psi(k,l,m)$).
The transformation is
\begin{equation}
\Psi(k,l,m)=k\int_{0}^{\pi}\sin\theta\,d\theta\int_{0}^{2\pi}d\varphi\,Y_{lm}^{*}(\theta,\varphi)\Psi_{\mathrm{0}}(\boldsymbol{k}),\label{eq:1}
\end{equation}
where $k=|\boldsymbol{k}|$ and $\hat{k}$ has spherical polar coordinates
$(\theta,\varphi)$.

To illustrate the method and avoid complications regarding wavepacket
spreading, we choose the simple, normalized momentum wavefunction
\begin{equation}
\Psi_{0}(\boldsymbol{k})=\frac{e^{-|\boldsymbol{k}-p\hat{\boldsymbol{z}}|^{2}/4\sigma_{p}^{2}}}{(2\pi\sigma_{p}^{2})^{\frac{3}{4}}}.\label{eq:2}
\end{equation}
The following calculation is simplest if the average momentum is chosen
in the $z$ direction. The standard deviation of each momentum component
is $\sigma_{p}.$

In a scattering experiment, we want the initial momentum to be well
resolved, so we choose
\begin{equation}
\epsilon\equiv\frac{\sigma_{p}}{p}\ll1.\label{eq:3}
\end{equation}
The results we derive below will be to lowest order in $\epsilon.$
It is the small magnitude of this parameter that will allow us to
construct an approximation method for which the leading term will
be sufficient for our purposes.

The spherical harmonic can be expressed in terms of a Wigner rotation
matrix as
\begin{equation}
Y_{lm}^{*}(\theta,\varphi)=\sqrt{\frac{2l+1}{4\pi}}e^{-im\varphi}d_{m0}^{l}(\theta),\label{eq:4}
\end{equation}
so
\begin{equation}
\int_{0}^{2\pi}d\varphi\,e^{-im\varphi}=2\pi\,\delta_{m0}.\label{eq:5}
\end{equation}

Then we use
\begin{equation}
|\boldsymbol{k}-p\hat{\boldsymbol{z}}|^{2}=(k-p)^{2}+2kp(1-\cos\theta).\label{eq:6}
\end{equation}
So the remaining integral for the wavefunction becomes
\begin{equation}
\Psi(k,l,m)=\delta_{m0}\sqrt{2\pi}\,k\,e^{-(k-p)^{2}/4\sigma_{p}^{2}}\sqrt{l+\frac{1}{2}}\int_{0}^{\pi}\sin\theta\,d\theta\,e^{-kp(1-\cos\theta)/2\sigma_{p}^{2}}d_{00}^{l}(\theta).\label{eq:7}
\end{equation}

It was intended to find an \textit{analytic} approximation to this
integral, so that we could make contact with results from partial
wave analysis and to minimize the amount of numerical computation
needed. The final calculation of the differential cross section then
requires only a numerical evaluation of a sum over $l$ \cite{Hoffmann2017a}.

In Eq. (\ref{eq:7}) the Gaussian in $k$ is only significant for
$k=p+\mathcal{O}(\sigma_{p})$. Then the exponential function of $\theta$
is sharply peaked at $\theta=0$ with a width of order $\epsilon.$
To evaluate this integral, we cannot use a Taylor series for the rotation
matrix, as for large $l$ it oscillates many times within the peak
of the exponential. Instead, we need an approximation valid for low
$\theta$ that is uniform in $l$. We use the method of Olver \cite{Olver1974}
for obtaining such expansions from the differential equation for the
function.

The Wigner rotation matrices are matrix elements of unitary rotations
about the $y$ axis,
\begin{equation}
d_{m_{1}m_{2}}^{j}(\theta)=\langle\,j,m_{1}\,|\,e^{-i\theta J_{y}}\,|\,j,m_{2}\,\rangle,\label{eq:a}
\end{equation}
and with the Condon-Shortley phase convention \cite{Messiah1961}
the matrix elements are all real.

The Wigner rotation matrices are predefined functions in M{\footnotesize{}ATHEMATICA}
(\texttt{WignerD}) \cite{Mathematica2017}. Note that the M{\footnotesize{}ATHEMATICA}
sign convention is
\begin{equation}
\mathtt{WignerD}[\{j,m_{1},m_{2}\},\theta]=d_{-m_{1},-m_{2}}^{j}(\theta).\label{eq:b}
\end{equation}

Note that in other cases of interest \cite{Jacob1959,Macfarlane1962},
more general rotation matrices $d_{m_{1}m_{2}}^{j}(\theta)$ will
appear in place of
\begin{equation}
d_{00}^{l}(\theta)=P_{l}(\cos\theta),\label{eq:8}
\end{equation}
including for half-integral angular momentum. We will derive a result
valid for the general case.

\section{Asymptotic approximation from the differential equation}

Wigner's series for the rotation matrix elements $d_{m_{1}m_{2}}^{j}(\theta)$
can be written in terms of a terminating hypergeometric series as
\cite{Rose1957} 
\begin{multline}
d_{m_{1}m_{2}}^{j}(\theta)=\left[\frac{(j+m_{1})!(j-m_{2})!}{(j-m_{1})!(j+m_{2})!}\right]^{\frac{1}{2}}\frac{(-)^{m_{1}-m_{2}}}{(m_{1}-m_{2})!}(\sin\frac{\theta}{2})^{m_{1}-m_{2}}(\cos\frac{\theta}{2})^{2j+m_{2}-m_{1}}\times\\
\times\phantom{|}_{2}F_{1}(-(j-m_{1}),-(j+m_{2});m_{1}-m_{2}+1;-\tan^{2}\frac{\theta}{2})\label{eq:9}
\end{multline}
for $m_{1}\geq m_{2}.$ We consider this regime first, then, for $m_{1}\leq m_{2},$
we use the symmetry relation \cite{Messiah1961}
\begin{equation}
d_{m_{1}m_{2}}^{j}(\theta)=(-)^{m_{1}-m_{2}}d_{m_{2}m_{1}}^{j}(\theta).\label{eq:10}
\end{equation}
This form gives us the small $\theta$ behaviour (again for $m_{1}\geq m_{2}$)
\begin{equation}
d_{m_{1}m_{2}}^{j}(\theta)\sim\left[\frac{(j+m_{1})!(j-m_{2})!}{(j-m_{1})!(j+m_{2})!}\right]^{\frac{1}{2}}\frac{(-)^{m_{1}-m_{2}}}{(m_{1}-m_{2})!}(\frac{\theta}{2})^{m_{1}-m_{2}},\label{eq:11}
\end{equation}
which we will use shortly.

An equivalent form is in terms of a Jacobi polynomial \cite{Szego1939}
\begin{multline}
d_{m_{1}m_{2}}^{j}(\theta)=(-)^{m_{1}-m_{2}}\left[\frac{(j+m_{1})!(j-m_{1})!}{(j+m_{2})!(j-m_{2})!}\right]^{\frac{1}{2}}(\sin\frac{\theta}{2})^{m_{1}-m_{2}}(\cos\frac{\theta}{2})^{m_{1}+m_{2}}\times\\
\times P_{j-m_{1}}^{(m_{1}-m_{2},m_{1}+m_{2})}(\cos\theta).\label{eq:12}
\end{multline}
The function
\begin{equation}
w(\theta)=(\sin\frac{\theta}{2}\cos\frac{\theta}{2})^{\frac{1}{2}}d_{m_{1}m_{2}}^{j}(\theta)\label{eq:13}
\end{equation}
obeys the particularly simple differential equation \cite{Abramowitz1972}
\begin{equation}
\{\frac{d^{2}}{d\theta^{2}}+(j+\frac{1}{2})^{2}-\frac{\alpha^{2}-\frac{1}{4}}{4\sin^{2}\frac{\theta}{2}}-\frac{\beta^{2}-\frac{1}{4}}{4\cos^{2}\frac{\theta}{2}}\}w=0,\label{eq:14}
\end{equation}
with
\begin{eqnarray}
\alpha & \equiv & m_{1}-m_{2},\label{eq:15}\\
\beta & \equiv & m_{1}+m_{2}.\label{eq:16}
\end{eqnarray}

Since we are looking for a low angle approximation, we expand the
trigonometric factors in powers of $\theta,$ keeping terms of order
$\theta^{2}$ in the differential equation. This gives the approximate
equation
\begin{equation}
\{\frac{d^{2}}{d\theta^{2}}+\Delta^{2}-\frac{\alpha^{2}-\frac{1}{4}}{\theta^{2}}+\psi(\theta)\}w=0,\label{eq:17}
\end{equation}
where
\begin{equation}
\Delta(j,m_{1},m_{2})\equiv\sqrt{j(j+1)-\frac{1}{3}(m_{1}^{2}+m_{2}^{2}+m_{1}m_{2}-1)}\label{eq:18}
\end{equation}
and
\begin{equation}
\psi(\theta)\sim\frac{(\alpha^{2}-\frac{1}{4})}{\Delta^{2}}\frac{\theta^{2}}{160}-\frac{(\beta^{2}-\frac{1}{4})}{\Delta^{2}}\frac{\theta^{2}}{16}\label{eq:19}
\end{equation}
for small $\theta.$ Note that
\begin{equation}
j(j+1)-\frac{1}{3}(m_{1}^{2}+m_{2}^{2}+m_{1}m_{2}-1)\geq j+\frac{1}{3}\label{eq:20}
\end{equation}
for given $j,$ so is always strictly positive.

Now we define
\begin{equation}
z\equiv\Delta\theta\label{eq:20.1}
\end{equation}
and use the transformation
\begin{equation}
w=\sqrt{z}\,y(z).\label{eq:20.2}
\end{equation}
Then the differential equation becomes
\begin{equation}
\{\frac{d^{2}}{dz^{2}}+\frac{1}{z}\frac{d}{dz}+1-\frac{\alpha^{2}}{z^{2}}+\psi(\frac{z}{\Delta})\}y=0.\label{eq:20.3}
\end{equation}
If the correction factor, $\psi(\theta),$ is sufficiently small and
can be neglected, this becomes the differential equation for the Bessel
function \cite{Gradsteyn1980} (the solution finite at the origin)
\begin{equation}
y(z)=J_{\alpha}(z).\label{eq:20.4}
\end{equation}
Instead of analytically calculating bounds on the error in our approximation,
we use numerical methods in Section III.

Now we have
\begin{equation}
w(\theta)\sim C\,\theta^{\frac{1}{2}}J_{m_{1}-m_{2}}(\Delta\theta),\label{eq:22}
\end{equation}
which then gives
\begin{equation}
d_{m_{1}m_{2}}^{j}(\theta)\sim D(\frac{\theta}{\sin\theta})^{\frac{1}{2}}J_{m_{1}-m_{2}}(\Delta\theta)\label{eq:23}
\end{equation}
for $m_{1}\geq m_{2}.$

To normalize the solutions, we note
\begin{equation}
J_{m_{1}-m_{2}}(\Delta\theta)\cong\frac{1}{(m_{1}-m_{2})!}(\Delta\frac{\theta}{2})^{m_{1}-m_{2}}\label{eq:25}
\end{equation}
for small $\theta.$ Comparing with Eq. (\ref{eq:11}), we find
\begin{equation}
D(j,m_{1},m_{2})=(-)^{m_{1}-m_{2}}\left[\frac{(j+m_{1})!(j-m_{2})!}{(j-m_{1})!(j+m_{2})!}\right]^{\frac{1}{2}}\frac{1}{\Delta^{m_{1}-m_{2}}}.\label{eq:26}
\end{equation}
Finally
\begin{multline}
d_{m_{1}m_{2}}^{j}(\theta)=(-)^{m_{1}-m_{2}}\left[\frac{(j+m_{1})!(j-m_{2})!}{(j-m_{1})!(j+m_{2})!}\right]^{\frac{1}{2}}\frac{1}{\Delta(j,m_{1},m_{2})^{m_{1}-m_{2}}}\times\\
\times(\frac{\theta}{\sin\theta})^{\frac{1}{2}}J_{m_{1}-m_{2}}(\Delta(j,m_{1},m_{2})\theta)+\mathcal{E}(j,m_{1},m_{2},\theta).\label{eq:27}
\end{multline}
for $m_{1}\geq m_{2}.$ We will find numerical bounds on the absolute
error, $|\mathcal{E}(j,m_{1},m_{2},\theta)|$, in the next section.
Note that
\begin{equation}
|d_{m_{1}m_{2}}^{j}(\theta)|\leq1\label{eq:28}
\end{equation}
from unitarity.

Note also that in typical applications we have
\begin{equation}
j\gg|m_{1}|,|m_{2}|,\label{eq:28.1}
\end{equation}
in which case
\begin{equation}
\left[\frac{(j+m_{1})!(j-m_{2})!}{(j-m_{1})!(j+m_{2})!}\right]^{\frac{1}{2}}\frac{1}{\Delta(j,m_{1},m_{2})^{m_{1}-m_{2}}}=1+\mathcal{O}(\frac{m_{i}}{j})\quad i=1,2.\label{eq:28.2}
\end{equation}

\section{Numerical calculation of error bounds}

In the applications we envision, for example the scattering of two
particles, $m_{2}$ will be a difference of helicities, not a large
number. For an impact parameter of, say, $10\,\sigma_{x}$, where
$\sigma_{x}=1/2\sigma_{p},$ we expect the wavefunction to only be
significant for $m_{1}\apprle10.$ Furthermore, for a typical choice,
$\epsilon=0.001$, the wavefunction will only be significant for $j\apprle2\,000.$

We first try a simple example that will be relevant to our original
problem, finding the error bound for
\begin{equation}
d_{00}^{j}(\theta)=P_{j}(\cos\theta)=(\frac{\theta}{\sin\theta})^{\frac{1}{2}}J_{0}(\sqrt{j(j+1)+\frac{1}{3}}\,\theta)+\mathcal{E}(j,\theta).\label{eq:29}
\end{equation}
We plot the absolute error, $|\mathcal{E}(j,\theta)|$, and $|P_{j}(\cos\theta)|$
(as calculated by M{\footnotesize{}ATHEMATICA}) against $\theta$
on log-log plots for (a) $j=10$ and (b) $j=2\,000$ in Figure 1.

As expected, we see the absolute and relative errors rising with angle.
For the small angles, $\theta\sim\epsilon$ that dominate the integral
Eq. (\ref{eq:7}), the relative error remains less than about $10^{-6}$
over the range of physical $j$ values. We plot the dependence on
$j$ for $\theta=0.001$ explicitly in Figure 2, confirming these
conclusions.

\begin{figure}
\begin{centering}
\includegraphics[width=16cm]{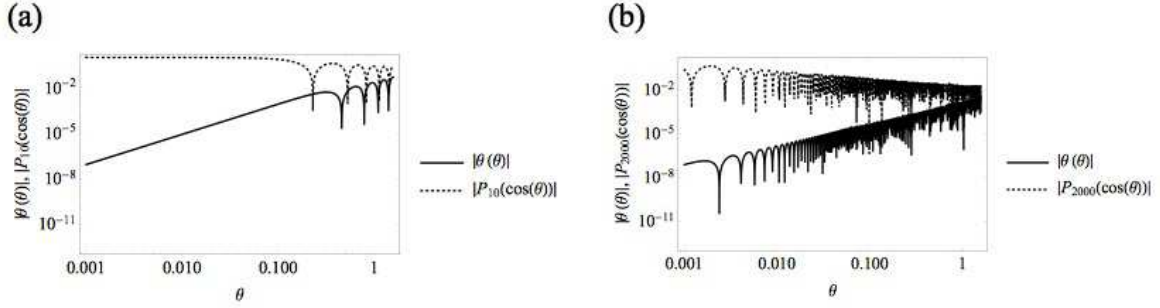}
\par\end{centering}
\caption{Absolute errors for our approximation of $d_{00}^{j}(\theta)$, compared
to $|P_{j}(\cos\theta)|$ for (a) $j=10$ and (b) $j=2\,000.$}

\end{figure}

\begin{figure}
\begin{centering}
\includegraphics[width=12cm]{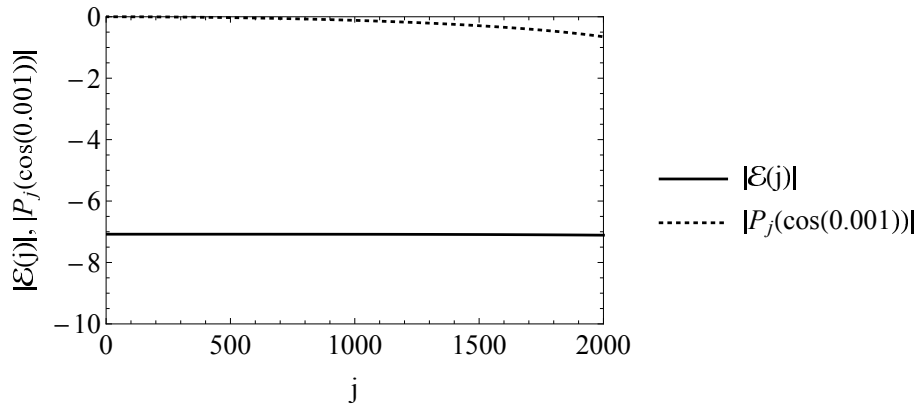}
\par\end{centering}
\caption{Absolute errors for our approximation of $d_{00}^{j}(0.001)$, compared
to $|P_{j}(\cos0.001)|$ for the physical range of $j.$}

\end{figure}

At another extreme, to see a case where our approximation may be less
valid, we investigate
\begin{equation}
d_{jj}^{j}(\theta)=(\frac{\theta}{\sin\theta})^{\frac{1}{2}}J_{0}(\sqrt{j+\frac{1}{3}}\,\theta)\{1+\mathcal{R}(j,\theta)\},\label{eq:33}
\end{equation}
using M{\footnotesize{}ATHEMATICA} for the exact Wigner matrices.
For this example we plot the \textit{relative} errors, defined by
\begin{equation}
\mathcal{R}(j,\theta)=\{d_{jj}^{j}(\theta)-(\frac{\theta}{\sin\theta})^{\frac{1}{2}}J_{0}(\sqrt{j+\frac{1}{3}}\,\theta)\}/d_{jj}^{j}(\theta),\label{eq:34}
\end{equation}
 in Figure 3. We see again that the error rises with angle, and it
also increases with $j.$ For the low angle $\theta=\epsilon$, the
relative error is less than $10^{-7}$ for $j=2000.$

\begin{figure}
\begin{centering}
\includegraphics[width=12cm]{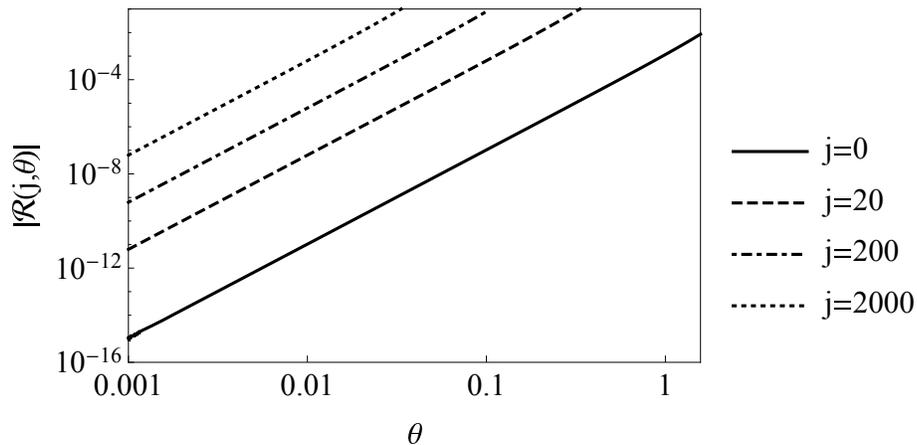}
\par\end{centering}
\caption{Relative errors for our approximation of $d_{jj}^{j}(\theta)$, for
$j=0,20,200,2000.$}

\end{figure}

We conclude with an example using half-integer spins,
\begin{equation}
d_{\frac{5}{2},\frac{1}{2}}^{j}(\theta)=\left[\frac{(j+\frac{5}{2})!(j-\frac{1}{2})!}{(j-\frac{5}{2})!(j+\frac{1}{2})!}\right]^{\frac{1}{2}}\frac{1}{j(j+1)-\frac{9}{4}}(\frac{\theta}{\sin\theta})^{\frac{1}{2}}J_{2}(\sqrt{j(j+1)-\frac{9}{4}}\,\theta)+\mathcal{E}(j,\theta)\label{eq:35}
\end{equation}
for $j\geq\frac{5}{2}$. The absolute error and $|d_{\frac{5}{2},\frac{1}{2}}^{j}(\theta)|$
(M{\footnotesize{}ATHEMATICA}) are plotted in Figure 4 for $j=2000.5$.
Again we find very small relative errors at low angles, rising with
angle.

\begin{figure}
\begin{centering}
\includegraphics[width=12cm]{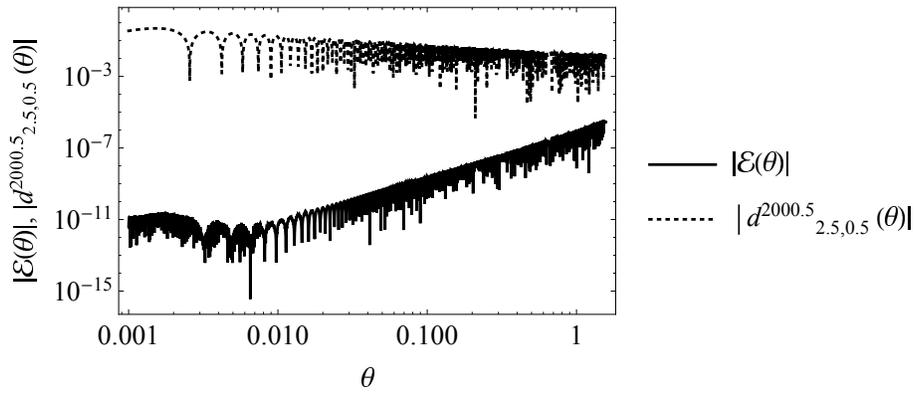}
\par\end{centering}
\caption{Absolute error for our approximation of $d_{\frac{5}{2},\frac{1}{2}}^{2000.5}(\theta)$.}

\end{figure}

\section{Approximation of the integral}

Returning to our original problem, we consider the factor from Eq.
(\ref{eq:7}) (with $\rho=k/p$)
\begin{equation}
I(\rho,l)=\frac{1}{2\epsilon^{2}}\int_{0}^{\pi}\sin\theta\,d\theta\,e^{-\rho(1-\cos\theta)/2\epsilon^{2}}P_{l}(\cos\theta).\label{eq:36}
\end{equation}
We make the further approximations $\sin\theta=\theta(1+\mathcal{O}(\theta^{2}))$
and $1-\cos\theta=\frac{\theta^{2}}{2}(1+\mathcal{O}(\theta^{2}))$
in the exponent and extend the upper limit of the integral to infinity
to find \cite{Gradsteyn1980}
\begin{eqnarray}
I(\rho,l) & \sim & \frac{1}{2\epsilon^{2}}\int_{0}^{\infty}\theta\,d\theta\,e^{-\rho\theta^{2}/4\epsilon^{2}}J_{0}(\sqrt{l(l+1)+\frac{1}{3}}\,\theta)\nonumber \\
 & = & \frac{1}{\rho}e^{-\epsilon^{2}(l(l+1)+\frac{1}{3})/\rho}\label{eq:39}
\end{eqnarray}
to lowest order in $\epsilon=\sigma_{p}/p$. A narrow distribution
in angle produces a wide distribution in angular momentum.

We define the relative error in this approximation as
\begin{equation}
\mathcal{R}(\rho,l)=\{I(\rho,l)-\frac{1}{\rho}e^{-\epsilon^{2}(l(l+1)+\frac{1}{3})/\rho}\}/I(\rho,l).\label{eq:40}
\end{equation}
In Figure 5 we plot the magnitude of this relative error for $\rho=1$
as a function of $l,$ up to three standard deviations. We see that
the relative error is $\apprle10^{-5}.$ The dependence on $\rho$
is very gradual, with the relative error changing by only $7\,\%$
of its $\rho=1$ value over $|\rho-1|\leq10\,\epsilon$ for $l=3000.$

\begin{figure}
\begin{centering}
\includegraphics[width=12cm]{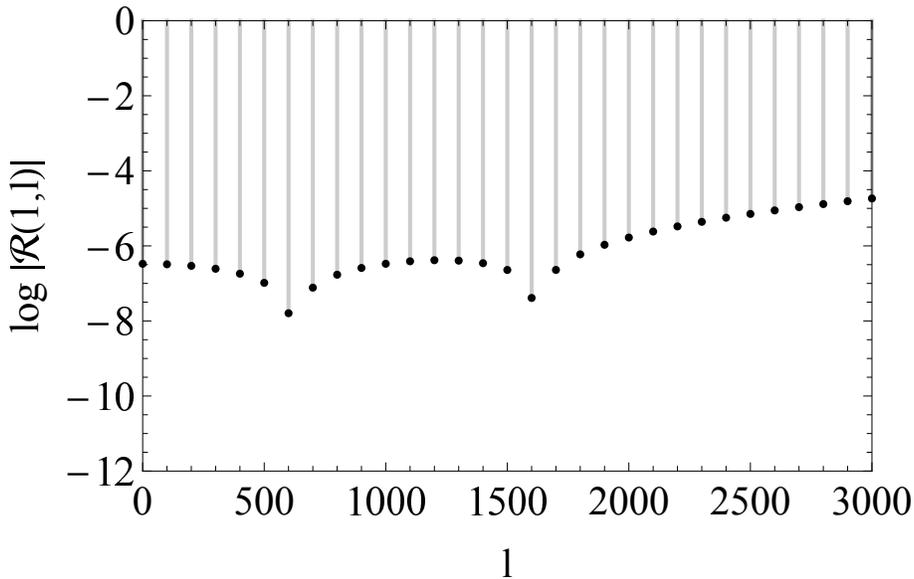}
\par\end{centering}
\caption{Relative error for $\rho=1.$}
\end{figure}

\section{Conclusions}

We have found a low angle approximation of the Wigner rotation matrix
elements, $d_{m_{1}m_{2}}^{j}(\theta)$, uniform in $j,m_{1}$ and
$m_{2}.$ Numerical determinations of errors in this approximation
have been given for a variety of cases. The relative error is reduced
if $j\gg|m_{1}|,|m_{2}|,$ which is the case in the applications we
envision. For our original problem of approximating a change of basis,
our method gives a relative error of $10^{-5}.$ We expect that this
approximation will have applications in the partial wave analysis
of wavepacket scattering.

The approximation presented here is merely the leading approximate
solution of a differential equation in the low angle region. It is
possible that an approximation with greater precision can be produced
by calculating additional terms.

\bibliographystyle{apsrev4-1}

\begin{thebibliography}{15}%
\makeatletter
\providecommand \@ifxundefined [1]{%
 \@ifx{#1\undefined}
}%
\providecommand \@ifnum [1]{%
 \ifnum #1\expandafter \@firstoftwo
 \else \expandafter \@secondoftwo
 \fi
}%
\providecommand \@ifx [1]{%
 \ifx #1\expandafter \@firstoftwo
 \else \expandafter \@secondoftwo
 \fi
}%
\providecommand \natexlab [1]{#1}%
\providecommand \enquote  [1]{``#1''}%
\providecommand \bibnamefont  [1]{#1}%
\providecommand \bibfnamefont [1]{#1}%
\providecommand \citenamefont [1]{#1}%
\providecommand \href@noop [0]{\@secondoftwo}%
\providecommand \href [0]{\begingroup \@sanitize@url \@href}%
\providecommand \@href[1]{\@@startlink{#1}\@@href}%
\providecommand \@@href[1]{\endgroup#1\@@endlink}%
\providecommand \@sanitize@url [0]{\catcode `\\12\catcode `\$12\catcode
  `\&12\catcode `\#12\catcode `\^12\catcode `\_12\catcode `\%12\relax}%
\providecommand \@@startlink[1]{}%
\providecommand \@@endlink[0]{}%
\providecommand \url  [0]{\begingroup\@sanitize@url \@url }%
\providecommand \@url [1]{\endgroup\@href {#1}{\urlprefix }}%
\providecommand \urlprefix  [0]{URL }%
\providecommand \Eprint [0]{\href }%
\providecommand \doibase [0]{http://dx.doi.org/}%
\providecommand \selectlanguage [0]{\@gobble}%
\providecommand \bibinfo  [0]{\@secondoftwo}%
\providecommand \bibfield  [0]{\@secondoftwo}%
\providecommand \translation [1]{[#1]}%
\providecommand \BibitemOpen [0]{}%
\providecommand \bibitemStop [0]{}%
\providecommand \bibitemNoStop [0]{.\EOS\space}%
\providecommand \EOS [0]{\spacefactor3000\relax}%
\providecommand \BibitemShut  [1]{\csname bibitem#1\endcsname}%
\let\auto@bib@innerbib\@empty
\bibitem [{\citenamefont {Dachsel}(2006)}]{Dachsel2006}%
  \BibitemOpen
  \bibfield  {author} {\bibinfo {author} {\bibfnamefont {H.}~\bibnamefont
  {Dachsel}},\ }\href@noop {} {\bibfield  {journal} {\bibinfo  {journal} {J.
  Chem. Phys.}\ }\textbf {\bibinfo {volume} {124}},\ \bibinfo {eid} {144115}
  (\bibinfo {year} {2006})}\BibitemShut {NoStop}%
\bibitem [{\citenamefont {Choi}\ \emph {et~al.}(1999)\citenamefont {Choi},
  \citenamefont {Ivanic}, \citenamefont {Gordon},\ and\ \citenamefont
  {Ruedenberg}}]{Choi1999}%
  \BibitemOpen
  \bibfield  {author} {\bibinfo {author} {\bibfnamefont {C.~H.}\ \bibnamefont
  {Choi}}, \bibinfo {author} {\bibfnamefont {J.}~\bibnamefont {Ivanic}},
  \bibinfo {author} {\bibfnamefont {M.~S.}\ \bibnamefont {Gordon}}, \ and\
  \bibinfo {author} {\bibfnamefont {K.}~\bibnamefont {Ruedenberg}},\
  }\href@noop {} {\bibfield  {journal} {\bibinfo  {journal} {J. Chem. Phys.}\
  }\textbf {\bibinfo {volume} {111}},\ \bibinfo {pages} {8825} (\bibinfo {year}
  {1999})}\BibitemShut {NoStop}%
\bibitem [{\citenamefont {Feng}\ \emph {et~al.}(2015)\citenamefont {Feng},
  \citenamefont {Wang}, \citenamefont {Yang},\ and\ \citenamefont
  {Jin}}]{Feng2015}%
  \BibitemOpen
  \bibfield  {author} {\bibinfo {author} {\bibfnamefont {X.~M.}\ \bibnamefont
  {Feng}}, \bibinfo {author} {\bibfnamefont {P.}~\bibnamefont {Wang}}, \bibinfo
  {author} {\bibfnamefont {W.}~\bibnamefont {Yang}}, \ and\ \bibinfo {author}
  {\bibfnamefont {G.~R.}\ \bibnamefont {Jin}},\ }\href {\doibase
  10.1103/PhysRevE.92.043307} {\bibfield  {journal} {\bibinfo  {journal} {Phys.
  Rev. E}\ }\textbf {\bibinfo {volume} {92}},\ \bibinfo {pages} {043307}
  (\bibinfo {year} {2015})}\BibitemShut {NoStop}%
\bibitem [{\citenamefont {Tajima}(2015)}]{Tajima2015}%
  \BibitemOpen
  \bibfield  {author} {\bibinfo {author} {\bibfnamefont {N.}~\bibnamefont
  {Tajima}},\ }\href {\doibase 10.1103/PhysRevC.91.014320} {\bibfield
  {journal} {\bibinfo  {journal} {Phys. Rev. C}\ }\textbf {\bibinfo {volume}
  {91}},\ \bibinfo {pages} {014320} (\bibinfo {year} {2015})}\BibitemShut
  {NoStop}%
\bibitem [{\citenamefont {Fukushima}(2016)}]{Fukushima2016}%
  \BibitemOpen
  \bibfield  {author} {\bibinfo {author} {\bibfnamefont {T.}~\bibnamefont
  {Fukushima}},\ }\href@noop {} {\emph {\bibinfo {title} {Numerical computation
  of {W}igner's d-function of arbitrary high degree and orders by extending
  exponent of floating point numbers}}},\ \bibinfo {type} {Tech. Rep.}\
  (\bibinfo  {institution} {National Astronomical Observatory of Japan},\
  \bibinfo {year} {2016})\BibitemShut {NoStop}%
\bibitem [{\citenamefont {Hoffmann}(2017)}]{Hoffmann2017a}%
  \BibitemOpen
  \bibfield  {author} {\bibinfo {author} {\bibfnamefont {S.~E.}\ \bibnamefont
  {Hoffmann}},\ }\href {http://stacks.iop.org/0953-4075/50/i=21/a=215302}
  {\bibfield  {journal} {\bibinfo  {journal} {J. Phys. B: At. Mol. Opt. Phys.}\
  }\textbf {\bibinfo {volume} {50}},\ \bibinfo {pages} {215302} (\bibinfo
  {year} {2017})}\BibitemShut {NoStop}%
\bibitem [{\citenamefont {Jacob}\ and\ \citenamefont {Wick}(1959)}]{Jacob1959}%
  \BibitemOpen
  \bibfield  {author} {\bibinfo {author} {\bibfnamefont {M.}~\bibnamefont
  {Jacob}}\ and\ \bibinfo {author} {\bibfnamefont {G.~C.}\ \bibnamefont
  {Wick}},\ }\href@noop {} {\bibfield  {journal} {\bibinfo  {journal} {Ann. of
  Phys. (N.Y.)}\ }\textbf {\bibinfo {volume} {7}},\ \bibinfo {pages} {404}
  (\bibinfo {year} {1959})}\BibitemShut {NoStop}%
\bibitem [{\citenamefont {Macfarlane}(1962)}]{Macfarlane1962}%
  \BibitemOpen
  \bibfield  {author} {\bibinfo {author} {\bibfnamefont {A.~J.}\ \bibnamefont
  {Macfarlane}},\ }\href@noop {} {\bibfield  {journal} {\bibinfo  {journal}
  {Rev. Mod. Phys.}\ }\textbf {\bibinfo {volume} {34}},\ \bibinfo {pages} {41}
  (\bibinfo {year} {1962})}\BibitemShut {NoStop}%
\bibitem [{\citenamefont {Olver}(1974)}]{Olver1974}%
  \BibitemOpen
  \bibfield  {author} {\bibinfo {author} {\bibfnamefont {F.~W.~J.}\
  \bibnamefont {Olver}},\ }\href@noop {} {\emph {\bibinfo {title} {Asymptotics
  and Special Functions}}}\ (\bibinfo  {publisher} {Academic Press, N.Y.},\
  \bibinfo {year} {1974})\BibitemShut {NoStop}%
\bibitem [{\citenamefont {Messiah}(1961)}]{Messiah1961}%
  \BibitemOpen
  \bibfield  {author} {\bibinfo {author} {\bibfnamefont {A.}~\bibnamefont
  {Messiah}},\ }\href@noop {} {\emph {\bibinfo {title} {Quantum Mechanics}}},\
  Vol.\ \bibinfo {volume} {1 and 2}\ (\bibinfo  {publisher} {North-Holland,
  Amsterdam and John Wiley and Sons, N.Y.},\ \bibinfo {year}
  {1961})\BibitemShut {NoStop}%
\bibitem [{Mat(2017)}]{Mathematica2017}%
  \BibitemOpen
  \href {https://www.wolfram.com} {\enquote {\bibinfo {title} {Wolfram
  {R}esearch {I}nc. {M}athematica},}\ }\bibinfo {howpublished}
  {https://www.wolfram.com} (\bibinfo {year} {2017})\BibitemShut {NoStop}%
\bibitem [{\citenamefont {Rose}(1957)}]{Rose1957}%
  \BibitemOpen
  \bibfield  {author} {\bibinfo {author} {\bibfnamefont {M.~E.}\ \bibnamefont
  {Rose}},\ }\href@noop {} {\emph {\bibinfo {title} {Elementary Theory of
  Angular Momentum}}}\ (\bibinfo  {publisher} {John Wiley and Sons, Inc., N.
  Y.},\ \bibinfo {year} {1957})\BibitemShut {NoStop}%
\bibitem [{\citenamefont {Szeg\"o}(1939)}]{Szego1939}%
  \BibitemOpen
  \bibfield  {author} {\bibinfo {author} {\bibfnamefont {G.}~\bibnamefont
  {Szeg\"o}},\ }\href@noop {} {\emph {\bibinfo {title} {Orthogonal
  Polynomials}}}\ (\bibinfo  {publisher} {A.M.S., Providence, R.I.},\ \bibinfo
  {year} {1939})\BibitemShut {NoStop}%
\bibitem [{\citenamefont {Abramowitz}\ and\ \citenamefont
  {Stegun}(1972)}]{Abramowitz1972}%
  \BibitemOpen
  \bibfield  {author} {\bibinfo {author} {\bibfnamefont {M.}~\bibnamefont
  {Abramowitz}}\ and\ \bibinfo {author} {\bibfnamefont {I.~A.}\ \bibnamefont
  {Stegun}},\ }\href@noop {} {\emph {\bibinfo {title} {Handbook of Mathematical
  Functions}}},\ \bibinfo {edition} {9th}\ ed.\ (\bibinfo  {publisher} {Dover,
  N. Y.},\ \bibinfo {year} {1972})\BibitemShut {NoStop}%
\bibitem [{\citenamefont {Gradsteyn}\ and\ \citenamefont
  {Ryzhik}(1980)}]{Gradsteyn1980}%
  \BibitemOpen
  \bibfield  {author} {\bibinfo {author} {\bibfnamefont {I.~S.}\ \bibnamefont
  {Gradsteyn}}\ and\ \bibinfo {author} {\bibfnamefont {I.~M.}\ \bibnamefont
  {Ryzhik}},\ }\href@noop {} {\emph {\bibinfo {title} {Tables of Integrals,
  Series and Products}}},\ \bibinfo {edition} {corrected and enlarged}\ ed.\
  (\bibinfo  {publisher} {Academic Press, Inc., San Diego, CA},\ \bibinfo
  {year} {1980})\BibitemShut {NoStop}%
\end{thebibliography}

\end{document}